\documentclass[aps,prl,superscriptaddress,showpacs,amsmath,twocolumn]{revtex4}

\usepackage{amsmath}
\usepackage{amssymb}
\usepackage{graphicx}

\begin{document}


\title{Unified Description of the Zitterbewegung
for Spintronic, Graphene, and Superconducting Systems}

\author{ J\'ozsef Cserti
}
\affiliation{Department of Physics of Complex Systems,
E{\"o}tv{\"o}s University
\\ H-1117 Budapest, P\'azm\'any P{\'e}ter s{\'e}t\'any 1/A, Hungary}
\author{Gyula D\'avid
}
\affiliation{Department of Atomic Physics,
E{\"o}tv{\"o}s University
\\ H-1117 Budapest, P\'azm\'any P{\'e}ter s{\'e}t\'any 1/A, Hungary}



\begin{abstract}

We present a unified treatment of Zitterbewegung phenomena for
a wide class of systems including spintronic, graphene, and
superconducting systems. We derive an explicit expression for the
time-dependence of the position operator of the quasiparticles which
can be decomposed into a mean part and an oscillatory term. The
latter corresponds to the Zitterbewegung. To apply our result
for different systems one needs to use only vector algebra instead
of the more complicated operator algebra.

\end{abstract}

\pacs{71.70.Ej, 73.63.Hs, 81.05.Uw, 73.43.Cd}

\maketitle


The {\em Zitterbewegung} (ZB) was first regarded as a relativistic effect
rooted in the Dirac equation and related to a `trembling' or
oscillatory motion of the center of a free wave
packet\cite{Schrodinger:cikk,Feschbach_Barut_Huang:cikk}. 
The ZB is caused by the interference between the positive
and negative energy states in the wave packet; the characteristic
frequency of this motion is determined by the gap between the two
states. It was believed that the experimental observation of the
effect is impossible since one would confine the electron to a scale
of the Compton wavelength $\hbar/m_0 c$, where $m_0$ is the bare
mass of the electron~\cite{Feschbach_Barut_Huang:cikk}. However, the
ZB is not a strictly relativistic effect: it can
appear even for a nonrelativistic particle moving in a
crystal~\cite{Cannata:cikkek} or for quasiparticles governed by the
Bogoliubov--de~Gennes equations in
superconductors~\cite{Lurie_Cremer:cikk}.

Most recently, Schliemann et
al.~\cite{Schliemann-1:cikk,D_Loss-PRB:cikk} predicted 
the ZB in spintronic systems where the experimental
observation of the effect is more realistic due to the much smaller
frequency of the oscillatory motion. In these semiconductor
nanostructures~\cite{spintronics-review:cikkek} spin-orbit coupling
generates an oscillatory motion of the wave packet. The
semi-classical time evolution of holes was investigated numerically
for the Luttinger Hamiltonian by Jiang et al.~\cite{Jiang:cikk}. The
relation between the ZB and the spin transverse
force was studied by Shen~\cite{Shen:cikk}. In a numerical work Lee
and Bruder observed an oscillatory behavior in the charge and spin
densities of quantum wires with Rashba and Dresselhaus types of
spin-orbit coupling~\cite{Lee_Bruder:cikk}. With a spin-polarized
electron injected into a waveguide, Nikoli\'c et
al.~\cite{Nikolic-SO_force:cikk} found an oscillatory motion of the
wave packet numerically, and the ZB pattern was
also predicted numerically by Brusheim and Xu~\cite{Brusheim:cikk}.
Similarly, Zawadzki studied the ZB in narrow gap
semiconductors~\cite{Zawadzki-1:cikk}, in single-wall semiconducting
carbon nanotubes~\cite{Zawadzki-2:cikk} and in crystals using the
nearly-free electron approximation~\cite{Zawadzki-3:cikk}, which is
essentially the same as the two-band model in~\cite{Cannata:cikkek}. 

Two-dimensional carbon sheets, known as graphene, have been studied
theoretically\cite{McClure_DiVincenzo:cikk,Graphene:cikkek} for many
decades, since their band structure is unique, a gapless Dirac-like
spectrum~\cite{Ziegler:cikkek}. However, the experimental
consequences of such a relativistic electron dynamics were
observed only recently in Hall
conductivity measurements~\cite{Novoselov-1:cikk,Kim:cikk}. 
In bilayer graphene a
more peculiar behavior of the Hall effect was observed
experimentally~\cite{Novoselov-2:cikk}, and explained in terms of
the chiral Hamiltonian first derived by McChann and
Fal'ko~\cite{Ed-Falko:cikk}. Both in single and bilayer graphene the
appearance of the oscillatory motion of the electron related to the
ZB was pointed out by
Katsnelson~\cite{Katsnelson:cikk}.
Most recently, Tworzyd\l o et al.\
associated the shot noise with the interference of electron-hole
pairs at the Dirac point in graphene~\cite{Tworzydlo:cikk}.
As an experimental observation of the ZB Trauzettel et al.\  proposed to
measure the photon-assisted electron transport 
in graphene~\cite{Trauzettel:cikk}. 

In this work we present a unified description of the ZB 
in the systems mentioned above. Our approach makes
it possible to calculate with simple algebra (without using operator
algebra) the time dependence of the position operator of the
particle for a wide class of systems. We also easily verify the
results first obtained by Schliemann et
al.~\cite{Schliemann-1:cikk,D_Loss-PRB:cikk} for spintronic systems.
Our result directly shows that the ZB is not
necessarily a relativistic effect but it is related to the coupling
between the components of the eigenstates of the system. This
phenomenon is thus the direct consequence of the pseudo-spin degree
of freedom.

The time-dependence of the position operator in the Heisenberg
picture is given by ${\bf r}(t) = e^{iHt/\hbar} \, {\bf r}(0)\,
e^{-iHt/\hbar}$, where $H$ is the Hamiltonian of the system. To
calculate the operator ${\bf r}(t)$ one can work with the
eigenstates of $H$. However, a further insight into the nature of
the ZB can be gained by solving the equations of
motion. We start with a quite general form of the Hamiltonian that
is suitable to describe the systems mentioned in the introduction:
\begin{equation}
H= \varepsilon({\bf p}) \openone
+ \mbox{\boldmath $\Omega$}^T\,
{\bf S},
\label{gen-H:eq}
\end{equation}
where the system is characterized by the one-particle energy
dispersion $\varepsilon({\bf p})$ and the effective magnetic field
$\mbox{\boldmath $\Omega$}({\bf p})$ coupled to the spin ${\bf S}$.
Here we assume that $\varepsilon({\bf p})$ and $\mbox{\boldmath
$\Omega$}({\bf p})$ are differentiable functions of the momentum
${\bf p}=(p_x,p_y,p_z)$. Here $T$ stands for the transpose of a
vector, while $\openone $ is the unit matrix in spin space, which
will be omitted hereafter. In the absence of an electrostatic
potential $V({\bf r})$ the momentum ${\bf p}$, and, consequently,
$\mbox{\boldmath $\Omega$}({\bf p})$, are constants of motion. In
Table~\ref{systems:table} we listed a few systems (together with the
effective magnetic field $\mbox{\boldmath $\Omega$}({\bf p})$) that
are currently intensely studied in spintronics, and in the research
of graphene and superconductors.

It should be emphasized that although in Table~\ref{systems:table}
the Hamiltonian for all systems is given in terms of the Pauli
matrices corresponding to a spin $S=\frac{1}{2}$, in our general
consideration, the spin operator ${\bf S}$ in Eq.~(\ref{gen-H:eq})
can represent a quasiparticle with an arbitrary spin $S \ge
\frac{1}{2}$.

\begin{widetext}
\begin{table*}[htb]
\begin{tabular}{|cccccc|} \hline \hline
system & $D$ & $H$ & $\mbox{\boldmath $\Omega$}$ & $\varepsilon({\bf p})$
& References   \\ \hline \hline
Rashba-Dresselhaus &
2 &
$ \begin{array}{c}
\frac{{\bf p}^2}{2m} +
\frac{\alpha}{\hbar}\, \left(p_x \sigma_y - p_y \sigma_x \right)
\\[1ex]
+ \frac{\beta}{\hbar}\, \left(p_y \sigma_y - p_x \sigma_x \right)
\end{array}$   &
$\frac{2}{\hbar^2} \left(
\begin{array}{c}
-\alpha \, p_y -\beta \, p_x \\
\alpha \, p_x +\beta \, p_y \\
0
\end{array}
\right)$ &
$\frac{{\bf p}^2}{2m}$   &
\cite{spintronics-review:cikkek,Schliemann-1:cikk,D_Loss-PRB:cikk,review_spinHall:cikkek}
\\ \hline
\begin{tabular}{c}
Heavy holes \\
in a quantum well
\end{tabular}
&
2 &
$\frac{{\bf p}^2}{2m} +  i \frac{\tilde{\alpha}}{2\hbar^3}
\left(p_-^3 \sigma_+ - p_+^3 \sigma_- \right)$    &
$\frac{2\tilde{\alpha}}{\hbar^4}
\left(
\begin{array}{cc}
p_y \left(3 p_x^2 - p_y^2 \right)  \\[1ex]
p_x \left(3 p_y^2 - p_x^2 \right) \\[1ex]
0
\end{array}
\right)$ &
$\frac{{\bf p}^2}{2m}$  &
\cite{Winkler_Schliemann-heavy-hole:cikkek,D_Loss-PRB:cikk,review_spinHall:cikkek}   \\ \hline
Bulk Dresselhaus &
3 &
$
\begin{array}{c}
\frac{\gamma_\textrm{D}}{\hbar^3}\left[
\sigma_x p_x\left(p_y^2 - p_z^2 \right)
+\sigma_y p_y\left(p_z^2 - p_x^2 \right) \right.
\\[1ex]
\left.
+\sigma_z p_z\left(p_x^2 - p_y^2 \right)
\right]
\end{array}
$ &
$\frac{2\gamma_{\textrm{D}}}{\hbar^4}
\left(
\begin{array}{c}
p_x \left(p_y^2 - p_z^2 \right)  \\[1ex]
p_y \left(p_z^2 - p_x^2 \right) \\[1ex]
p_z \left(p_x^2 - p_y^2 \right)
\end{array}
\right)$  &
$0$  &
\cite{spintronics-review:cikkek,review_spinHall:cikkek}  \\ \hline
\begin{tabular}{c}
Single-layer \\
graphene
\end{tabular}  &
2 &
$v \left(p_x \sigma_x + p_y \sigma_y \right)$ &
$\frac{2v}{\hbar}\, \left(
\begin{array}{cc}
p_x  \\
p_y  \\
0
\end{array}
\right)$ &
0  &
\cite{McClure_DiVincenzo:cikk,Novoselov-1:cikk,Kim:cikk,Graphene:cikkek,Katsnelson:cikk}  \\ \hline
Bilayer graphene &
2 &
$ \frac{1}{2m} \left(\frac{p_+^2 +p_-^2 }{2}\sigma_x
-\frac{p_-^2 - p_+^2 }{2i}\sigma_y\right)$ &
$\frac{1}{m\hbar} \left(
\begin{array}{c}
p_x^2 - p_y^2 \\
2 p_x p_y \\
0
\end{array}
\right)$  &
0   &
\cite{Novoselov-2:cikk,Ed-Falko:cikk,Katsnelson:cikk} \\ \hline
Cooper pairs  &
3 &
$\left(\frac{{\bf p}^2}{2m}-E_F \right) \sigma_z + \Delta \sigma_x$ &
$\frac{2}{\hbar} \left(
\begin{array}{c}
\Delta  \\
0 \\
\frac{p_x^2 + p_y^2 + p_z^2}{2m} - E_F
\end{array}
\right)$  &
0 &
\cite{BdG:book,Lurie_Cremer:cikk} \\ \hline
Nearly free electrons &
3 &
$H= \left(\begin{array}{cc}
\epsilon_{{\bf k}+{\bf q}} & V_{\bf q} \\[1ex]
V^*_{\bf q} & \epsilon_{{\bf k}}
\end{array}
\right)$
&
$\left(
\begin{array}{cc}
\Re\{V_{\bf q}\}  \\[1ex]
-\Im\{V_{\bf q}\} \\[1ex]
\frac{1}{2}\, (\epsilon_{{\bf k}+{\bf q}} -\epsilon_{{\bf k}})
\end{array}
\right)$
&
$\frac{1}{2}\, (\epsilon_{{\bf k}+{\bf q}} +\epsilon_{{\bf k}})$ &
\cite{Zawadzki-3:cikk,Cannata:cikkek}
\\ \hline \hline
\end{tabular}
\caption{The Hamiltonian of different systems can be expressed as
in Eq.~(\ref{gen-H:eq}).
Here $D$ is the dimension of the system,
$p_\pm = p_x \pm i p_y$, $\sigma_\pm = \sigma_x  \pm i
\sigma_y$, and the spin operator is
${\bf S}= \frac{\hbar}{2}\, \mbox{\boldmath $\sigma$}$, where
$\mbox{\boldmath $\sigma$}=(\sigma_x,\sigma_y,\sigma_z)$ is the set of Pauli
matrices.
For Cooper pairs we assume (for simplicity) that the pair potential
$\Delta$ is real and independent of ${\bf r}$,
and that the energy is measured from the Fermi energy $E_F$.
In the last row,
$\epsilon_{\bf p} =  \hbar^2 {\bf k}^2/(2m)$, where ${\bf p}=\hbar{\bf k}$,
${\bf q}$ is fixed, and
$V_{\bf q}$ is the Fourier transform of the periodic potential
treated as a perturbation in the crystal.
Here $\Re\{\cdot \}$ and $\Im\{\cdot \}$ are the real
and imaginary parts of the argument.
More details of these systems can be found in the references listed
in the last column.
\label{systems:table}}
\end{table*}
\end{widetext}

The equations of motion of the position operator ${\bf r}(t)$
and the spin operator ${\bf S}(t)$
in the Heisenberg picture for the Hamiltonian (\ref{gen-H:eq}) read
\begin{subequations}
\begin{eqnarray}
\frac{d}{dt}\,{\bf r}(t) &=& \frac{i}{\hbar} \left[ H,{\bf r} \right] =
\frac{d\varepsilon({\bf p})}{d {\bf p}}
+ {\bf K}\, {\bf S}(t),
\label{r_motion:eq} \\
\frac{d}{dt}\, {\bf S}(t)
&=& \frac{i}{\hbar} \left[ H,{\bf S} \right] =
\mbox{\boldmath $\Omega$}({\bf p}) \times {\bf S}(t), \, \text{where}
\label{sigma_motion:eq}  \\
K_{ik} &=& -\frac{i}{\hbar}\,\left[x_i,\Omega_k({\bf p})\right]
=  \frac{\partial \Omega_k}{\partial p_i}.
\label{K:def}
\end{eqnarray}%
\label{r_sigma_motion:eq}%
\end{subequations}%
Note that Eqs.~(\ref{r_sigma_motion:eq}) are coupled equations of all
three components of $ {\bf S}(t)$ and ${\bf r}(t)$.
However, in the case of two-dimensional systems only the $x$- and
$y$-components of ${\bf r}(t)$ are involved in Eq.~(\ref{r_motion:eq}).

It is clear from Eq.~(\ref{sigma_motion:eq}) that the spin vector
${\bf S}(t)$ precesses around the vector
$\mbox{\boldmath $\Omega$}$.
The solution of Eq.~(\ref{sigma_motion:eq}) with the initial
condition ${\bf S}(0)= {\bf S}_0$
can be written as
\begin{equation}
{\bf S} (t) = \left[{\bf n}\circ{\bf n}
+ \left(\openone - {\bf n}\circ{\bf n}\right) \cos \Omega t
+ \sin\Omega t \,\, {\bf n} \times \right]  {\bf S}_0,
\label{sigma_t_sub:eq}
\end{equation}
where $\mbox{\boldmath $\Omega$} = \Omega \, {\bf n}$, ${\bf n}$
is a unit vector,
$\Omega^2 = \mbox{\boldmath $\Omega$}^T \mbox{\boldmath $\Omega$}$,
and ${\bf n}\circ{\bf n}$ denotes
the outer or direct product, i.e.,
${({\bf n}\circ{\bf n})}_{ik} = n_i n_k$.
Here the operator ${\bf S}_0$ on the right-hand side is
in the Schr\"odinger picture, i.e., it is time independent.
One can show that the usual commutation relations still hold
$\left[S_i(t),S_j(t)\right] = i\hbar \varepsilon_{ijk} S_k(t)$.

Inserting Eqs.~(\ref{sigma_t_sub:eq}) into Eq.~(\ref{r_motion:eq})
and solving the differential equation one finds
\begin{eqnarray}
{\bf r}(t)
&=& {\bf r}_0
+ \frac{1}{\Omega}\, {\bf K} \left({\bf n} \times {\bf S}_0 \right)
+\frac{d\varepsilon({\bf p})}{d {\bf p}} \, t
+ ({\bf K n}) ({\bf n S}_0) \, t  \nonumber \\[1ex]
&& \hspace{-7mm}  +\frac{\sin \Omega t }{\Omega} \,
{\bf K} \left(I-{\bf n}\circ{\bf n} \right)\, {\bf S}_0
- \, \frac{\cos \Omega t }{\Omega}\,
{\bf K} \left({\bf n} \times {\bf S}_0 \right),
\label{rH_t:eq}
\end{eqnarray}%
with the initial condition ${\bf r}(0)={\bf r}_0$.
This is our central result.
The interpretation of the different terms in (\ref{rH_t:eq}) is as
follows.
The ZB stems from the oscillatory
terms (cosine and sine terms).
In contrast to the usual dynamics (first and third terms),
two new terms appear in the non-oscillatory part:
the transverse displacement, which is
independent of time (second term), and a term that corresponds to
a particle motion with constant anomalous velocity (third term).
In addition to the oscillatory part, these two terms
in ${\bf r}(t)$ are inherent of the ZB.
The anomalous velocity plays a crucial role in the anomalous and spin
Hall effects in semiconductors~\cite{similar_H:cikkek}.

To evaluate the time-dependent position operator ${\bf r}(t)$ within a
Gaussian wave packet one can follow, e.g., the calculation presented
in Refs.~\cite{Schliemann-1:cikk,D_Loss-PRB:cikk}.

We are now in a position to apply our results to the systems listed in
Table~\ref{systems:table}.
Using Eqs.~(\ref{K:def}) and (\ref{rH_t:eq}), some simple algebra yields
the same results as given by Eqs.~(7) and (8) in
Ref.~\cite{D_Loss-PRB:cikk} for the Rashba-Dresselhaus system.
Similarly, it is easy to verify the results Eqs.~(41) and (42) in
Ref.~\cite{D_Loss-PRB:cikk} for systems of heavy holes
in a quantum well~\cite{note:sajat}.

The current operator in graphene systems splits into three terms
of which the last one can be associated with the
ZB phenomenon~\cite{Katsnelson:cikk}.
Our general approach can also be applied to graphene layers to find
the time evolution of the position operator ${\bf r}(t)$.
We now present explicit results for the position operator ${\bf r}(t)$ from
which the trembling (oscillatory) motion of the electron in graphene
systems is clearly seen.
For single-layer and bilayer graphene, the explicit formulas
for $x(t)$ and $y(t)$ can again be easily obtained
using Table~\ref{systems:table} and Eqs.~(\ref{K:def}) and (\ref{rH_t:eq}).
The results for single-layer graphene are
\begin{subequations}
\begin{eqnarray}
x(t) &=& x_0+ v\sigma_x t
+ \frac{p_y}{p^2}\,\frac{\hbar}{2}\, \sigma_z
\left[1- \cos\left(\frac{2pv}{\hbar}\,t \right)\right]
\nonumber \\[1ex]
 + && \hspace{-7mm} \frac{p_y}{p^3} \frac{\hbar}{2}
\left(p_x \sigma_y - p_y \sigma_x \right)
\left[\frac{2pv}{\hbar}\,t -\sin\left(\frac{2pv}{\hbar}\,t \right)
\!
\right]
\! , \\[1ex]
y(t) &=& y_0+ v\sigma_y t
- \frac{p_x}{p^2}\,\frac{\hbar}{2}\, \sigma_z
\left[1- \cos\left(\frac{2pv}{\hbar}\,t \right)
\right]
\nonumber \\[1ex]
 - && \hspace{-7mm}
\frac{p_x}{p^3} \frac{\hbar}{2}
\left(p_x \sigma_y - p_y \sigma_x \right)
\left[\frac{2pv}{\hbar}\,t -\sin\left(\frac{2pv}{\hbar}\,t \right)
\!
\right]
\! ,
\end{eqnarray}%
\end{subequations}%
and for bilayer graphene
\begin{subequations}
\begin{eqnarray}
x(t) &=& x_0 + \frac{p_x \sigma_x +p_y \sigma_y }{m}\, t
+\frac{p_y}{p^2}\, \hbar\, \sigma_z
\left( 1- \cos \Omega t \right)
\nonumber \\[1ex]
&& -
\frac{ p_y \, \Sigma_G }{p^4} \,
\left( \Omega t - \sin \Omega t \right)
, \\[1ex]
y(t) &=& y_0 + \frac{-p_y \sigma_x +p_x \sigma_y }{m}\, t
-\frac{p_x}{p^2}\, \hbar\, \sigma_z
\left( 1- \cos \Omega t \right)
\nonumber \\[1ex]
&& +
\frac{ p_x \, \Sigma_G }{p^4}
\left( \Omega t - \sin \Omega t \right), \\[1ex]
\Sigma_G &=&
\hbar \left[2p_x p_y \sigma_x - (p_x^2-p_y^2)\sigma_y \right],
\hspace{3mm} \Omega = \frac{p^2}{\hbar m} .
\end{eqnarray}%
\end{subequations}%
Here $\sigma_i$ are the Pauli matrices and $p^2 = p_x^2 + p_y^2$.

Similarly, using the Hamiltonian for Cooper pairs given
in Table~\ref{systems:table} the following results are obtained:
\begin{subequations}
\begin{eqnarray}
{\bf r}(t) &=& {\bf r}_0 +  \frac{{\bf p}}{m}\, \sigma_x t
+ \frac{{\bf p}}{m} \, \Sigma_C \left( \Omega t -\sin \Omega t \right)
\nonumber \\[1ex]
&+& \frac{{\bf p}}{m}\, \frac{\Delta}{E^2(p)} \frac{\hbar}{2}\, \sigma_y
\left( 1- \cos \Omega t \right), \\[1ex]
E(p) &=& \sqrt{{\left(\frac{p^2}{2m} - E_F \right)}^2 + \Delta^2},
\hspace{3mm}
\Omega = \frac{2 E(p)}{\hbar} , \\[1ex]
\Sigma_C &=&  \frac{\hbar}{2} \frac{\Delta}{E^3(p)}
\left[\left(\frac{p^2}{2m}-E_F\right)\, \sigma_x - \Delta \sigma_z\right].
\end{eqnarray}%
\end{subequations}%
Here $p^2 = p_x^2 + p_y^2 + p_z^2$.
One can show that these results agree with those presented
in Ref.~\cite{Lurie_Cremer:cikk}.

Similarly, some simple algebra yields the same results as in
Refs.~\cite{Cannata:cikkek,Zawadzki-3:cikk} for nearly free electrons listed
in Table~\ref{systems:table} (except that the off-diagonal elements
are swapped in the latter reference).
For bulk Dresselhaus systems (3rd row in Table~\ref{systems:table})
the calculation is again straightforward but the results are rather
cumbersome and not presented here.

{\em Discussion.}
As mentioned above, the position operator ${\bf r}(t)$ in
(\ref{rH_t:eq}) is decomposed into a mean part and an oscillatory term.
If one derives the position operator ${\bf r}(t)$ directly from
${\bf r}(t) = e^{iHt/\hbar} \, {\bf r}(0)\, e^{-iHt/\hbar}$ then
such a decomposition can only be obtained using the
Foldy-Wouthuysen transformation~\cite{Foldy:cikk,Lurie_Cremer:cikk}.
In this transformation the operator ${\bf r}(t)$ is calculated in the
basis of the eigenstates of Hamiltonian (\ref{gen-H:eq}).
It can be shown that for $S=\frac{1}{2}$ (with Pauli matrices)
the eigenenergies $E_\pm ({\bf k}) $ and the
eigenstates $\psi_\pm({\bf r}) = |\chi_\pm \rangle e^{i {\bf k r}}$
are given by
\begin{subequations}
\begin{eqnarray}
E_\pm ({\bf k}) &=&  \varepsilon(\hbar{\bf k})
\pm \frac{\hbar}{2} |\mbox{\boldmath $\Omega$}(\hbar{\bf k})|, \\[1ex]
|\chi_+ \rangle &=& \left(
\begin{array}{c}
\cos \frac{\Theta}{2} e^{-i\frac{\Phi}{2}}  \\[1ex]
\sin \frac{\Theta}{2} e^{i\frac{\Phi}{2}}
\end{array}
\right) ,
|\chi_- \rangle = \left(
\begin{array}{c}
-\sin \frac{\Theta}{2} e^{-i\frac{\Phi}{2}}  \\[1ex]
\cos \frac{\Theta}{2} e^{i\frac{\Phi}{2}}
\end{array}
\right),
\end{eqnarray}
\end{subequations}
where $\Theta$ and $\Phi$ are the spherical polar angles of the vector
$\mbox{\boldmath $\Omega$}(\hbar{\bf k})$ in ${\bf k}$-space,
and $|{\bf a}|$ is the magnitude of vector ${\bf a}$.
However, for $S> \frac{1}{2}$ the Foldy-Wouthuysen transformation is more
complicated.
The advantage of our approach is that it leads directly to the
desired decomposition of the position operator ${\bf r}(t)$.

For pure Rashba coupling and for single-layer graphene
the ZB can be interpreted as a consequence of
the conservation of the total angular momentum $J_z= L_z + S_z$,
where ${\bf L}= {\bf r} \times {\bf p}$ is the orbital angular
momentum (see Ref.~\cite{D_Loss-PRB:cikk}).
However, in general, $J_z$ is not a constant of motion, ie.,
$\left[H, {\bf J}\right] =
\mbox{\boldmath $\Omega$} \times {\bf S}
- {\bf p}  \times {\bf K} {\bf S} \neq 0$.
As it can be readily seen, this is the case, for example,
for Rashba-Dresselhaus systems where $\alpha \neq 0$ and $\beta \neq 0$,
or for heavy holes in a quantum well.

Finally, it should be mentioned that the ZB is related to
the non-trivial behavior of the conductivity of
single and bilayer graphenes~\cite{Katsnelson:cikk} since
the velocity operator~(\ref{r_motion:eq}) does not commute with
the Hamiltonian (\ref{gen-H:eq}).
The peculiar behavior of the spin Hall effect may also be related to
the ZB~\cite{Shen:cikk}.

We gratefully acknowledge discussions
with C. W. J. Beenakker, J. Schliemann, B. Nikoli\'c,
V. Fal'ko, T. Geszti, and A. Pir\'oth.
This work is partly supported by E.~C.\ Contract No.~MRTN-CT-2003-504574.



\begin{thebibliography}{10}
\bibitem{note:sajat}
After consulting the authors of Ref.~\cite{D_Loss-PRB:cikk} we
clarified a misprint in their paper:
the factors $p_x$ and $p_y$ in the 4th terms of Eqs.~(41) and (42),
respectively, should be swapped.
\end{thebibliography}
\end{document}